\documentclass[%
12pt,
 reprint,
 amsmath,amssymb,
 aps,
]{revtex4-2}

\usepackage{graphicx}
\usepackage{dcolumn}
\usepackage{bm}

\begin{document}


\title{\LARGE Spin-orbit Interaction-mediated Measurement of Surface Chirality}

\author{\large Upasana Baishya}
\altaffiliation[ ]{School of Physics, University of Hyderabad, Hyderabad 500046, India.}

\author{\large Nirmal K. Viswanathan}%
\altaffiliation[ ]{School of Physics, University of Hyderabad, Hyderabad 500046, India.}
\email{nirmalsp@uohyd.ac.in}


\begin{abstract}
\textbf{The spin-orbit ($\sigma-l$) interaction in a focused-reflected beam of light results in spatially non-uniform polarization in the beam cross-section due to the superposition of orthogonal field components and polarization-dependent interface reflection coefficients. Polarization filtering the output beam leads to an interchangeable transformation of $l = \mp 2$ charge vortex into two ($\mp$) unit charge vortices, for $\sigma = \pm 1$ circular polarization of the input Gaussian beam. This transformation follows a trajectory, named optical vortex trajectory (OVT), that depend on the input beam’s $\sigma$ and hence the $l$ and reflecting surface characteristics. The OVT is used here to quantify both the sign and the magnitude of the chiral parameter of a quartz crystal. The Jones matrix-based simulation anticipates the chirality-dependent OVT that matches with experimental measurements. }
\end{abstract}

\maketitle


Understanding the reflection-refraction of electromagnetic waves at a dielectric interface is one of the fundamental quests of optical physics \cite{jackson2021classical,born19997th}. The relative amplitude of reflection and refraction of polarized light at the interface between two transparent, homogenous, isotropic, achiral media was first deduced by Fresnel \cite{born19997th}. Independent of this, he also explained optical rotation of linear polarized light passing through an optically active medium as a manifestation of circular birefringence, arising due to the difference in phase velocities of orthogonal circular polarization states of light \cite{born19997th,silverman2008chiral}. Despite these seminal contributions, it was not until the mid-1980 that a formal understanding of light reflection from a homogeneous, isotropic, transparent, and absorbing chiral media was realized by Silvermann and his co-workers \cite{silverman2008chiral,silverman1985specular,silverman1986effects,silverman1986reflection,silverman1988experimental,silverman1992chiral}.

Manifestation of chirality, associated with light reflection from a transparent chiral medium, includes (i) polarization-dependent inequality of angles of reflection, (ii) generation of elliptically polarized light from incident linearly polarized light, and (iii) differential reflection of right and left circularly polarized light \cite {silverman1986effects,silverman1988experimental}. In the electrodynamic treatment, the interaction of the electric field with a chiral medium is described by $D =\epsilon[E-i\gamma(k×E)]$, where the real parameter $\gamma$, designated as gyrotropic or chiral parameter characterizes the interaction strength \cite{silverman1986effects,silverman1990light}. For natural optically active medium (such as z-cut quartz), the refractive index for right (+) and left (-) circularly polarized light is given by $n_\pm^2=\epsilon \mu (1 \pm \gamma)$ and the ellipticity of the reflected beam, for linear (TE or TM) polarized incident beam, is given by $e_{TE}=e_{TM}=\gamma[n⁄(n^2-1)$]. The chiral parameter $\gamma$=0 for inactive dielectric medium (such as glass). The small value of the chiral parameter ($\gamma \sim 0.1$) makes the experimental measurement of its manifestation challenging, unless under special conditions \cite{silverman1989large,silverman1989differential,silverman1992multiple,ghosh2006chiral}.

It was shown recently by our research group that reflection of an optical beam at an air-dielectric interface gives rise to significant polarization non-uniformity or inhomogeneity in the beam cross-section \cite{debnath2020generalized,debnath2021generic,debnath2022generic,debnath2023optical,kumar2024complex}. The polarization inhomogeneity is engineered to observe optical singularities, and related beam-field phenomena. These are due to the spin-orbit interaction (SOI) of light including the spin-to-orbit conversion (SOC) and the spin-Hall effect of light (SHEL) in a generalized context. In the special case of normal incidence and reflection of a converging or diverging beam, the angle of incidence (AoI) of the central wavevector $\theta_{i0}=0^\circ$, while the AoI of the surrounding constituent wavevectors (or rays) are non-zero. Interesting phase - polarization structures and trajectories are simulated and measured as a function of the input beam polarization ellipticity variation. Notably, an on-axis $\pm 2$ charge  optical vortex (OV) splits into two off-centred $\pm 1$ charge vortices \cite{debnath2023optical}. The appearance of OVs and their dynamics depend on the input beam polarization and the refractive index of the reflecting surface. We use here reflection from a chiral surface to measure the modification to the polarization inhomogeneity and the vortex trajectory and use them to quantify the chiral parameter. The use of OVs combined with the weak measurement scheme allows us to extract the small chiral parameter very accurately. 

\begin{figure}[htbp]
\centering
\includegraphics[scale=0.14]{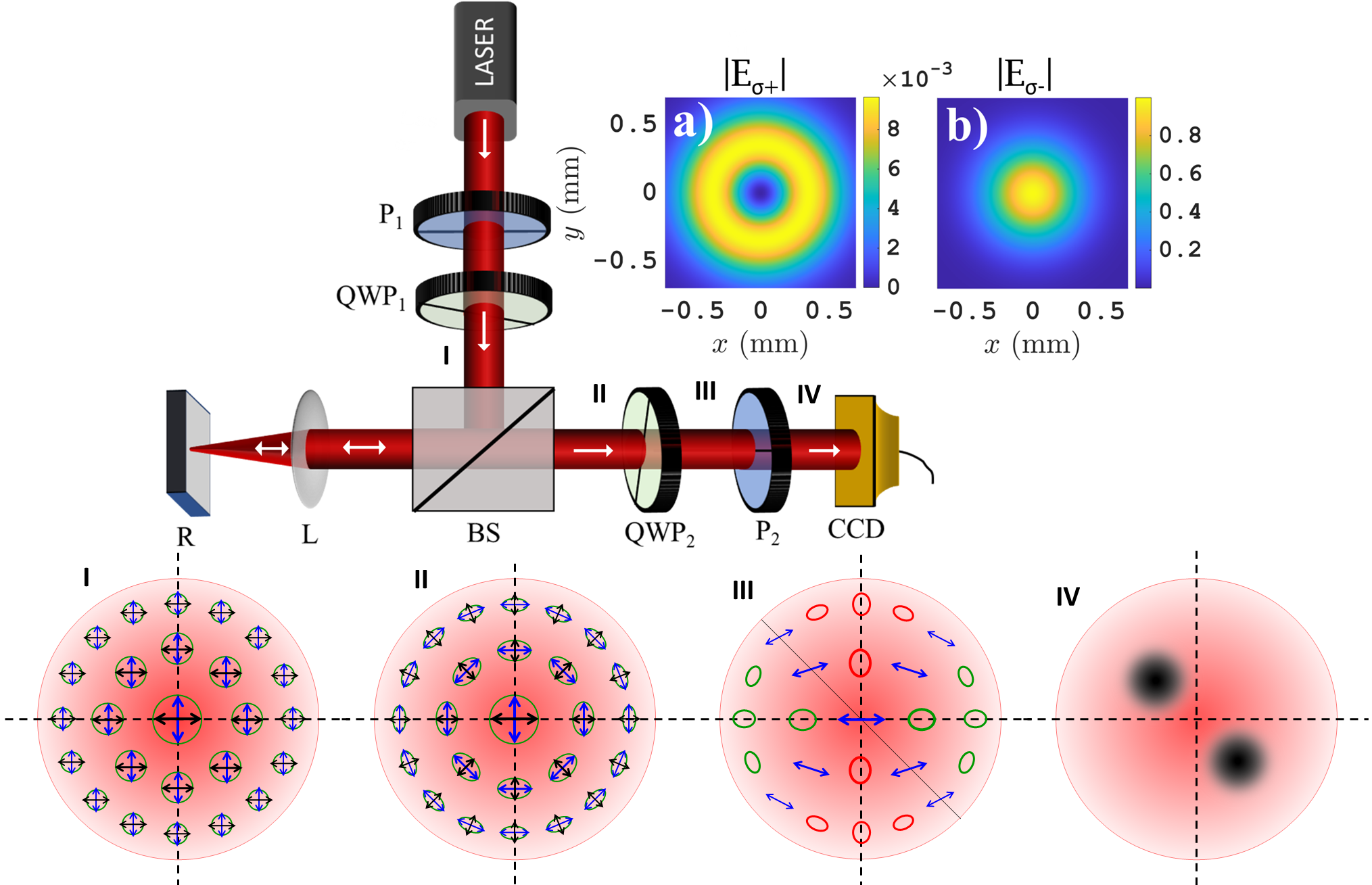}
\caption{\label{fig1} Schematic of the optical setup used to study the vortex trajectory arising due to the reflection of a focused light beam at an air-chiral interface. R- reflecting surface, L- high NA lens, BS- 50-50 non-polarizing beam splitter, $QWP_1$, $QWP_2$- quarter-wave plates, $P_1$, $P_2$- polarizers and CCD- camera. Inset (a) and (b) show the simulated field components $E_{\sigma^+}$ and $E_{\sigma^-}$ at the back focal plane of L for $\sigma^-$ input Gaussian beam. The SoP of the reflected beam is visualised at different positions of the setup for ${\sigma^-}$ polarised input light (I). (II) spatially inhomogeneous left elliptically polarised light, (III) combination of linear, left, and right elliptically polarised light after $QWP_2$ (oriented at $45^\circ$), (IV) beam intensity after $P_2$ (oriented at $-\theta$).}
\vspace{2pt}
\hrule
\end{figure}

Schematic of the optical system is shown in Fig.\ref{fig1}. The Gaussian beam from a laser is circularly polarised using a polarizer ($P_1$) and quarter-wave plate ($QWP_1$). The beam is then reflected at a non-polarising 50-50 beam-splitter (BS) and tightly focused using a 100X microscope objective lens L (DIN Semi-Plan, Edmund). A partially reflecting surface is introduced at the focal plane of the lens so that the reflected light gets collimated upon retro-reflection. The collimated output beam is then transmitted through the same BS and its characteristics are measured using a CCD camera. A quarter-wave plate ($QWP_2$) and polarizer ($P_2$) combination kept in front of the CCD camera enables measurement of the state of polarization (SoP) of the output beam via Stokes polarimetry and also to carry out weak measurement to study the characteristics of the reflecting surface. For ease of visualisation, the exaggerated view of te SoP of the reflected beam at different positions (II, III, IV) are shown in Fig.\ref{fig1} for left-circular polarized (LCP, $\sigma^-$) input Gaussian beam (I).

We use Richards and Wolf treatment for focused Gaussian beam and the vectorial angular spectrum method to calculate the electric field components at the focal plane for LCP ($\hat{\sigma}^-$) input Gaussian beam \cite{richards1959electromagnetic,novotny2012principles}:
\begin{align} \label{e1}
    &E^{\hat{\sigma}^-}_f=\displaystyle\int^{\theta_{max}}_0 A(\theta,Z) \begin{bmatrix}
    (1-cos\theta)\; j_2(x) \; e^{-2i\psi} \; \\
     (1+cos\theta)\; j_0(x)\; \\
      2i\; sin\theta \; j_1(x)\; e^{-i\psi}
    \end{bmatrix}
    \begin{bmatrix}
       {\hat{\sigma}^+} \\ {\hat{\sigma}^-} \\ \hat{z}
    \end{bmatrix}
\end{align}
Here, $ A(\theta,Z) = \cfrac{\sqrt{2}\:i\:\pi}{\lambda} f \: \sqrt{cos\theta}\:sin\theta\:e^{-ik\,Z\,cos\theta}$ is the amplitude factor. With $x =k \rho sin\theta$, $ j_0(x),j_1(x)$ and $j_2(x)$ are the Bessel functions of first kind of order 0,1 and 2, respectively. $\theta$ is the focusing related cone angle made by the wave vectors of light beam with the propagation axis and $\psi$ is the azimuth angle. It is clear from Eq.\ref{e1}, that at the focus, the input LCP beam has decomposed into RCP, LCP, and Z components of vortex charge -2, 0 and -1, respectively due to focusing, and attributed to the SOI of light \cite{debnath2023optical}.

A partially reflecting dielectric surface kept at the focal plane, and perpendicular to the central wave vector $\Vec{\mathrm{k}_\circ}$ (corresponding to $0^\circ$ AoI) reflects the beam back through the same lens. Except for $\Vec{\mathrm{k}_\circ}$, all other wavevectors in the focused beam exhibit variation in the SoP to maintain the transversality condition. Combining this SoP variation with the radially varying and azimuthally symmetric reflection coefficients $\mathrm{\begin{bmatrix}r_{pp} & r_{ps} \\ r_{sp} & r_{ss} \end{bmatrix}}$ due to the change in the AoI, results in inhomogeneously polarised reflected beam in the output \cite{debnath2023optical}. As shown in Fig.\ref{fig1}, focusing and retro-reflecting LCP input Gaussian beam (I) results in nonuniform SoP at the output (II) that depends on the reflecting surface characteristics. The field components of the output beam, after reflection at the air-dielectric interface, in circular polarization basis is
\begin{align}\label{e2}
     \begin{bmatrix}
        \boldsymbol{E_{\sigma^+}} \\ \boldsymbol{E_{\sigma^-}}
    \end{bmatrix}
      = -{\cfrac{B(\rho)}{\sqrt{2}}} \;  \begin{bmatrix}
   { (r_{pp} + r_{sp} + r_{ss} + r_{ps})\: \; e^{- 2\:i\:\phi}\; \; } \\
    { (r_{pp} + r_{sp} - r_{ss} - r_{ps})\: } 
    \end{bmatrix}   
    \begin{bmatrix}
        \hat{\sigma}^+ \\ \hat{\sigma}^-
    \end{bmatrix}
 \end{align} 
Here ${B(\rho) = \cfrac{E_o}{\sqrt{2}}\; exp(-\cfrac{\rho^2}{\omega_0^2}) \;exp(2 i k z_0)}$, $\rho$ is the radial distance from $\Vec{\mathrm{k}_\circ}$ and ${\omega_0}$ is the beam waist. The reflection coefficients are given by \cite{lekner1996optical} 

\begin{align*}\label{e3}
r_{pp} = - \cfrac{a - b}{c + d}\; ; \;\; r_{ss} = \cfrac{a + b}{c + d}\; ; \;\; r_{ps} =r_{sp}= \cfrac{e}{c + d}
\end{align*}
\begin{align*}
    & a=(1-g^2)(cos\theta_+ + cos\theta_-)cos\theta ,\\
    & b=2g(cos^2\theta - cos\theta_+cos\theta_- \\
    & c=(1+g^2)(cos\theta_+ + cos\theta_-)cos\theta ,\\
    & d=2g(cos^2\theta + cos\theta_+cos\theta_-) \\
    & e= -2ig(cos\theta_+ - cos\theta_-)cos\theta
\end{align*}

Where ${g = \sqrt{\epsilon/\mu} \; , \; cos\theta_\pm = \sqrt{1-\cfrac{sin^2\theta}{n^2_\pm}}}$ and ${ n_\pm = (n \pm \gamma)}$, and $n_+$ and $n_-$ are the refractive indices for left and right circular polarized light. The off-diagonal elements $r_{ps}, r_{sp}$ are zero for optically inactive dielectric material and $\gamma$ is the chiral parameter tensor of optically active crystal \cite{condon1937theories,tellegen1948gyrator}. For isotropic material, $\gamma$ is single-valued. The sample studied here is a z-cut, R-quartz crystal of point group 32 with $\gamma$ =$\gamma_{11}, \gamma_{33} \sim \gamma \times 10^{-6}$. In our system, instead of a single AoI we access a range $ \sim [0^\circ,65^\circ]$, corresponding to the NA of the lens, in a single beam cross-section. So the $\gamma$ parameters we refer to is $((\gamma_{11}+\gamma_{33})/2)$ and the sign of the effective $\gamma$ will indicate the R/L characteristic of the quartz.

We consider LCP ($\sigma^-$) Gaussian input light beam of wavelength $\lambda$ = 632.8 nm with power $P_0$ = 1 mW, focused by a 100x microscope objective lens with NA $\sim 1 $. The focused beam is reflected at the air-glass ($\gamma = 0$) interface of refractive indices 1.0 and 1.457. The orthogonal RCP ($\sigma^+$) field component, in the retro-reflected, collimated beam is a Laguerre-Gaussian (LG) beam with a vortex of charge $l$ = -2 (Fig.\ref{fig1}a). The field strength of the $\sigma^+$ component is two orders of magnitude smaller than the dominant $\sigma^-$ term (Fig.\ref{fig1}b). The sign of the vortex charge depends on the SAM ($\sigma = \pm1$) of the input Gaussian beam. The input beam's SAM gets partially converted into orbital angular momentum (OAM) in the output due to SOC.
\begin{figure}[htbp]
\centering
\includegraphics[scale=0.057]{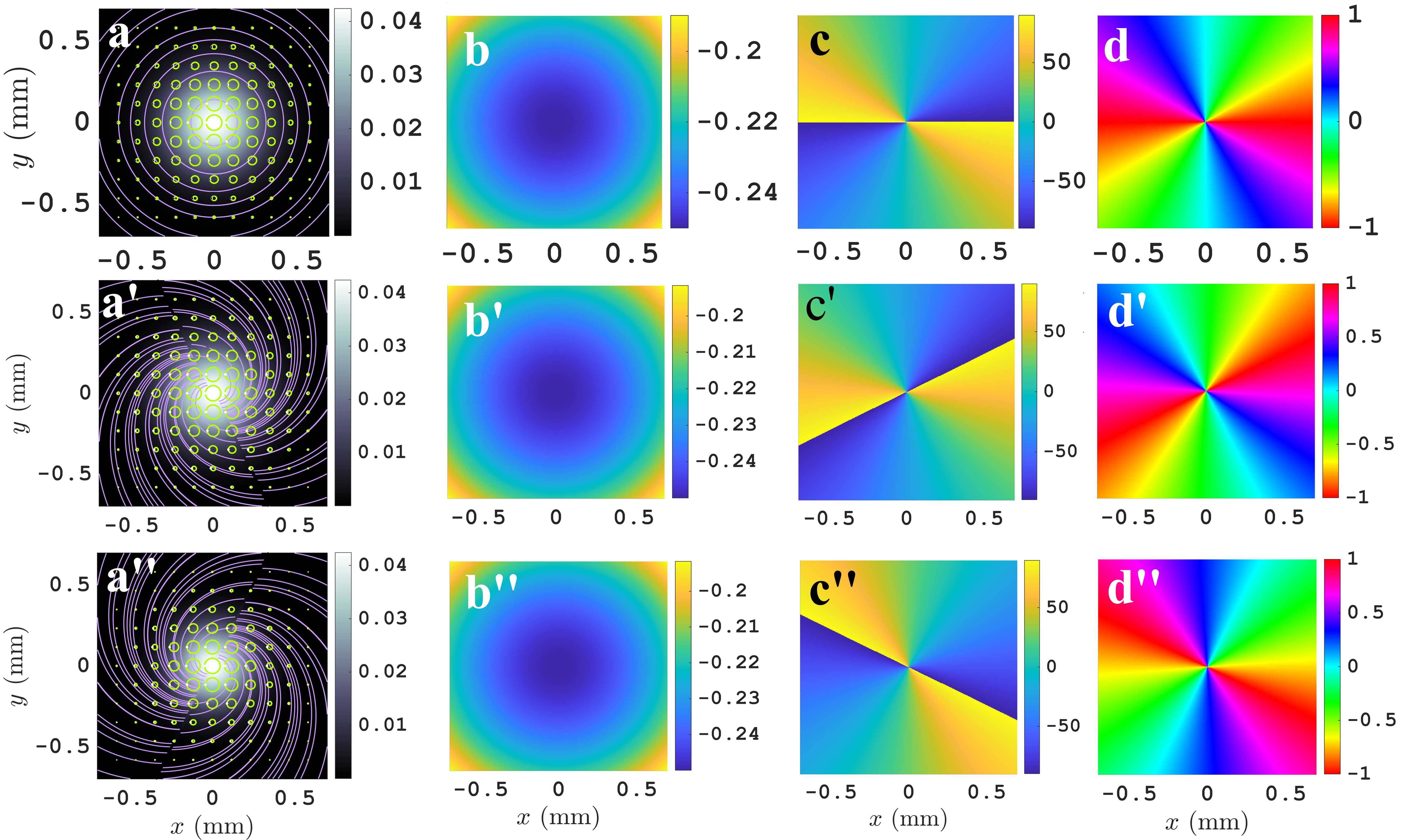}
\caption{\label{fig2} (a) Simulated image of focused-reflected-collimated output beam polarization with streamlines, (b) spatial variation of the polarization ellipticity, (c) ellipse orientation, and (d) phase difference. First row is for reflection at air-glass interface and row 2 and 3 are due to reflection at right- and left- chiral interface respectively.} 
\vspace{2pt}
\hrule
\end{figure}

Since both field components ($\sigma = \pm1$ with $l = - 2, 0$) are present simultaneously in the reflected beam, their coherent superposition results in spatially inhomogeneous field distribution. However, the two orders of magnitude difference in their intensity results in weak spatial variation of the SoP as seen in Fig.\ref{fig1} II. Figure \ref{fig2}(a) - (d)  shows the spatial variation of the SoP in the output beam cross-section with streamlines, the polarization ellipticity $\chi$, polarization ellipse orientation $\Psi$ and the phase difference $\phi_{12}$ between right- and left- circular polarization for air-glass reflecting interface. These are calculated respectively using the Stokes parameters:

\begin{center} 
$sin\:2\chi = \cfrac{S_3}{S_0}$ \hspace{0.1cm} , \hspace{0.1cm} $tan\:2\Psi = \cfrac{S_2}{S_1}$ \hspace{0.1cm} and \hspace{0.1cm} 
$\phi_{12} = angle(S_1+i\:S_2)$
\end{center}

Here ${S_0, S_1, S_2, S_3}$ are the output beam Stokes parameters \cite{goldstein2017polarized}. The spatial variation in the SoP is due to the cone angle of the focused beam and corresponding Fresnel reflection coefficients \cite{baishya2023measurement}. It is important to point out here that it is the variations in the polarization parameters which allows us to measure the surface reflectivity characteristics of the interface. Each point in the beam cross-section represents superposition of different values of the ($\sigma^- -\sigma^+$) field components, resulting in different amounts of polarization ellipticity and ellipse orientation in the beam cross-section. The phase difference in circular basis (Fig.\ref{fig2}d) shows a $4 \pi$ variation corresponding to -2 charge OV at the center. Also, the beam center (Fig.\ref{fig2}a) with undefined polarization ellipse orientation corresponds to C-point polarization singularity. Next we calculate the phase and polarization characteristics of the output beam for right- and left- handed (quartz) chiral sample. As can be seen (Fig.\ref{fig2} (a'-d') and Fig.\ref{fig2} (a" - d")), the streamlines, the polarization ellipse orientation and the phase difference for the chiral samples rotate counter clockwise (CCW) or clockwise (CW) for right or left chiral quartz respectively. This clearly emphasises the variation that arises due to the SOI of the focused beam of light with the chiral sample. The CW and CCW rotation of the SoP streamlines around the C-point singularity is akin to the spiral topological structure and  the Airy spiral due to propagation of light through a chiral medium \cite{volyar2007optical}.

\begin{figure}[htbp]
\centering
\includegraphics[scale=0.05]{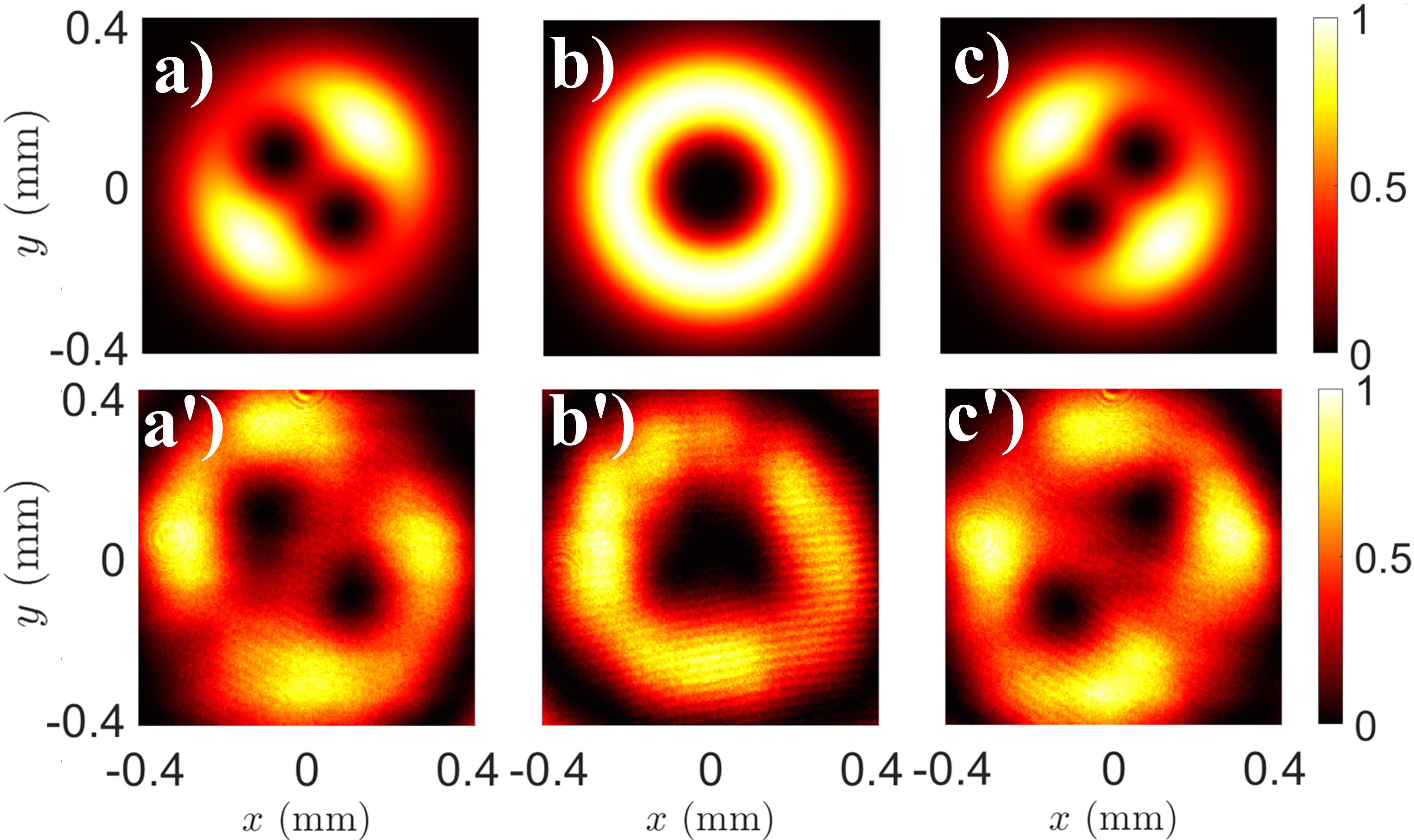}
\caption{\label{fig3} (a) - (c) Simulated and (a') - (c') experimentally measured images due to reflection at air-chiral interface. (b), (b') correspond to the cross-polarization condition, (a), (a') and (c), (c') correspond to $\pm 1^\circ$ angle of the analyzer.}
\vspace{2pt}
\hrule
\end{figure}

We now use the Jones matrix formalism to calculate the final intensity, after cross-polarization filtering using a quarter wave plate ($QWP_2$) and a polariser ($P_2$) combination. Taking the Jones matrix for the QWP and the polariser as $J_{q2}$ and $J_{p2}$ respectively, the output field components at the Fourier plane is calculated using \cite{chipman2018polarized},
\begin{equation} \label{e3}
      \begin{bmatrix}
        E^{'}_{\sigma+}\\ E^{'}_{\sigma-}   
    \end{bmatrix} = 
    J_{p2}.\;J_{q2}.
    \begin{bmatrix}
        E_{{\sigma+}}\\E_{{\sigma-}}   
    \end{bmatrix}
\end{equation}

Where, $E_{{\sigma+}}$ and $E_{{\sigma-}}$ denote the respective field components of the light beam after the BS (Eq.\ref{e2}). The output beam intensity is given by 
        $ I = E{^{'\:2}_{\sigma+}} + E{^{'\:2}_{\sigma-}} $.
        
With the $QWP_2 = +45^\circ$ and analyzer $P_2 = 90^\circ$, the output beam intensity corresponding to orthogonal circular polarisation state to the input ($P_1 = 0^\circ, \; QWP_1 = -45^\circ$) is a vortex beam with $-2$ charge  (Fig.\ref{fig3}b). The optical vortex (OV) in the orthogonal polarization is a manifestation of spin-to-orbit conversion (SOC) of tightly focused and reflected light beam \cite{debnath2023optical}. Fixing the $QWP_2$ orientation, if we rotate $P_2$ by a small amount, the -2 charge OV splits into two -1 charge OVs. The position of the two OVs in the beam cross-section depends on the analyzer angle. As we rotate the analyzer angle more, the two OVs move away from each other along a linear trajectory, and eventually move completely out of the beam cross-section  for the analyzer angle ($\gtrsim 3^\circ$). Rotating $P_2$ in the opposite direction, starting from the cross position, results in the two OVs splitting and moving along a linear path, perpendicular to the previous trajectory. Figure \ref{fig3}a and \ref{fig3}c show the intensity distribution for the two analyzer angle of $\mp \; 1^\circ $. The behavior indicates the spatially inhomogeneous nature of the SoP in the beam cross-section and can be understood based on the mechanism as illustrated in Fig.\ref{fig1} at III and IV positions.

\begin{figure}[htbp]
\centering
\includegraphics[scale=0.058]{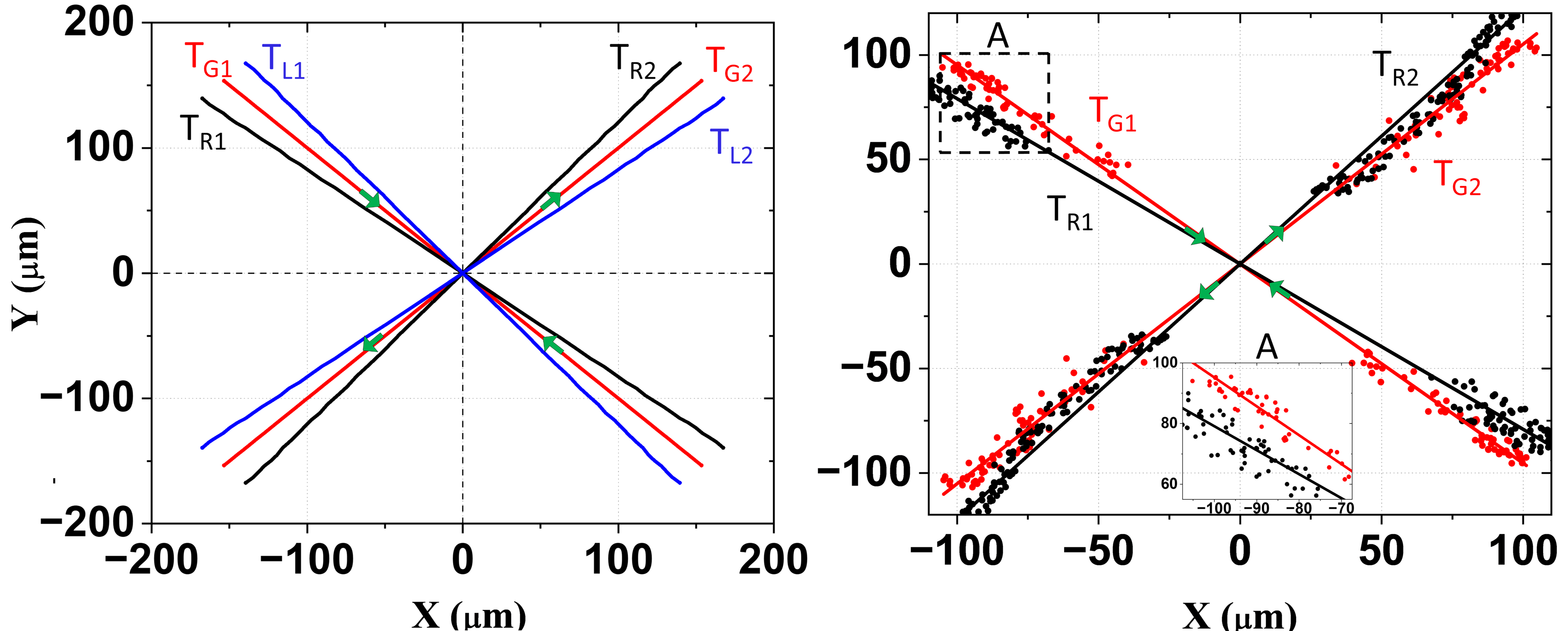}
\caption{\label{fig5} (a) Simulated OVT as a function of chiral parameter $\gamma$: red ($T_{G1}-T_{G2}$) for air-glass interface ($\gamma$=0), black ($T_{R1}-T_{R2}$) and blue ($T_{L1}-T_{L2}$) for air-chiral ($\gamma= \pm 0.15$) interface reflection. (b) Experimentally measured OVT for refection at air-glass (red symbol), and air-R-quartz (black symbol) interface. Black and red colour lines are corresponding linear fit. Inset A shows a zoomed view of the trajectories.}
\vspace{2pt}
\hrule
\end{figure}

Depending on the reflection coefficient of the dielectric interface, the inhomogeneous SoP of the output beam and, hence, the position of the optical vortices will be different for the same analyzer angle range. Tracking the vortex position as a function of the analyser angle we get the optical vortex trajectory (OVT), which is used here to differentiate between glass and chiral reflection, between right- and left chiral reflection and for $\sigma^- -\sigma^+$ input polarization. Figure \ref{fig5}a is a plot of the OVT in the x-y plane of the beam as a function of analyzer angle. Trajectories $T_{G1}$ and $T_{G2}$ (red colour lines) correspond to air-glass interface and trajectories $T_{R1} - T_{R2}$ (black lines) and $T_{L1} - T_{L2}$ (blue lines) respectively correspond to the right and left chiral quartz surface reflections. For the parameters used here, all the vortex trajectories start in the 2-4 quadrant and end in the 1-3 quadrant for CW rotation of the analyser angle from $-\theta$ to $+\theta$. The chiral parameter of the glass and of the right and left chiral (quartz) surfaces used for the simulation are respectively $\gamma$ = 0, +0.15 and -0.15. For all trajectories, starting from the analyser angle $-2^\circ$ the vortex positions are tracked. The two -1 charge OVs approach each other, and at the orthogonal position ($P_2$=$90^\circ$), merge to become a single -2 charge OV at the beam center. Continuing to rotate the analyzer angle beyond this point, the two vortices appear, start separating and move along a trajectory orthogonal to the first trajectory. The end positions of the vortices correspond to the analyzer angle $+2^\circ$. The direction of arrows on the trajectory indicates the path of the optical vortices. 

As can be seen, with reference to $\gamma$= 0, the OVT of R-quartz and L-quartz are distinctly different. The slope for each pair of straight line corresponding to different $\gamma$ are calculated. For glass ($\gamma$= 0), the value of slopes $T_{G1}$ and $T_{G2}$ are $\mp 1$ and the angle both the trajectory lines make with the x-axis is $45^\circ$. For $\gamma$= +0.15, the value of slopes of $T_{R1}, T_{R2}$ are $-0.828$, $+1.208$ and the angle, the  trajectory lines make with x-axis are $39.61^o$ and $50.39^o$ respectively . The OST of L-quartz with  $\gamma$= -0.15, the values of slopes $T_{L1}$ and $T_{L2}$ are $-1.208^\circ$ and $+0.828^\circ$, respectively. It is important to note here that changing the $\gamma$ value changes the reflection coefficients, which leads to a change in the SoP of the output beam and hence the OVT for the same analyser angle range. Also, with an increase/decrease of the $\gamma$ value, the position of the OVs, the trajectory slope and the angle made will all be different for the same analyzer angles.

Finally we demonstrate the anticipated effects arising due to the SOI of light by reflecting a focused beam of light at air-dielectric (glass or chiral) interface. Schematic of the experimental setup shown in Fig.\ref{fig1}. In the experimental demonstration, the thickness of the sample is taken to be larger than the Rayleigh range of the lens, to avoid the back surface reflected light affecting the front surface reflection. We first calibrate the experimental setup using an air-glass interface. The output beam with -2 charge vortex (Fig.\ref{fig3}b') is obtained for the orthogonal position of $QWP_2$ = $45^\circ$ and $P_2$ = $90^\circ$. Rotating $P_2$ by $-\theta$ = $-1^\circ$ and $+1^\circ$ around this angle we observe splitting of the -2 charge vortex into to two -1 charge off-axis vortices in the beam cross-section Figure \ref{fig3}a' and \ref{fig3}c'. show the separation of the vortices corresponding to these analyzer angles respectively. The output beam intensity, measured using the CCD camera, as a function of the analyzer angle is then used to calculate the OVT. The experimentally measured OVT for air-glass (red) and air-R-quartz (black) interface reflections as a function of $P_2$ angle are shown in (Fig.\ref{fig5}b). The  continuous line are the corresponding linear fit. The value of the two slopes $T_{G1}$ and $T_{G2}$ for glass are $-0.953 \pm 0.005$ and $1.052 \pm 0.007$ respectively. The angle made by the trajectory lines with the x-axis are $43.62^o$ and $46.44^o$ respectively. The OVT for R-quartz is rotated CCW with reference to the glass trajectory. The value of the two slopes $T_{R1}$ and $T_{R2}$ for R-quartz are $-0.791 \pm 0.003$ and $1.223 \pm 0.003$, respectively. The angles made by the trajectory lines with the x-axis are $38.36^o$ and $50.74^o$, respectively. From the measured experimental data, the chiral parameter $\gamma$ of the R-quartz due to surface reflection is calculated to be $+0.15\pm 0.005$, matching with the theoretical calculation. Though the experiment is not carried out for the L-quartz (due to its non-availability) we anticipate a CW rotation of the OVT, matching with the behaviour shown in (Fig.\ref{fig5}a). The chiral parameter sign indicates the chiral material's (R/L) handedness.

We demonstrated a novel method, of using OVT arising due to the SOI, to study the surface chirality leading to the measurement of chiral parameter ($\gamma$), and handedness of a chiral material. By retro-reflecting a focused beam of light at an air–dielectric interface, we measure the OVT and use it to quantify $\gamma$ of the chiral surface. As the reflected beams' SoP is sensitive to the azimuthally symmetric and radially varying surface reflection coefficients, the resulting change in the OVT is a direct measure of the surface chirality. The sensitivity and accuracy of the method can be improved further in experimental implementation and data analysis. The proposed and demonstrated method is highly suitable to characterize chirality of ultra-thin film and of 2D material, either in reflection or in transmission.

\vspace{1cm}
\textbf{\large  Funding:}  Science and Engineering Research Board (SERB) and Institution of Eminence (IoE), University of Hyderabad.

\textbf{\large Acknowledgments:} The authors acknowledge SERB for financial support to this area of research and Institution of Eminence (IoE), University of Hyderabad for research fellowship (UB).

\textbf{\large Disclosures:} The authors declare no conflicts of interest.

\bibliography{sample1}

\begin{thebibliography}{27}%
\makeatletter
\providecommand \@ifxundefined [1]{%
 \@ifx{#1\undefined}
}%
\providecommand \@ifnum [1]{%
 \ifnum #1\expandafter \@firstoftwo
 \else \expandafter \@secondoftwo
 \fi
}%
\providecommand \@ifx [1]{%
 \ifx #1\expandafter \@firstoftwo
 \else \expandafter \@secondoftwo
 \fi
}%
\providecommand \natexlab [1]{#1}%
\providecommand \enquote  [1]{``#1''}%
\providecommand \bibnamefont  [1]{#1}%
\providecommand \bibfnamefont [1]{#1}%
\providecommand \citenamefont [1]{#1}%
\providecommand \href@noop [0]{\@secondoftwo}%
\providecommand \href [0]{\begingroup \@sanitize@url \@href}%
\providecommand \@href[1]{\@@startlink{#1}\@@href}%
\providecommand \@@href[1]{\endgroup#1\@@endlink}%
\providecommand \@sanitize@url [0]{\catcode `\\12\catcode `\$12\catcode `\&12\catcode `\#12\catcode `\^12\catcode `\_12\catcode `\%12\relax}%
\providecommand \@@startlink[1]{}%
\providecommand \@@endlink[0]{}%
\providecommand \url  [0]{\begingroup\@sanitize@url \@url }%
\providecommand \@url [1]{\endgroup\@href {#1}{\urlprefix }}%
\providecommand \urlprefix  [0]{URL }%
\providecommand \Eprint [0]{\href }%
\providecommand \doibase [0]{https://doi.org/}%
\providecommand \selectlanguage [0]{\@gobble}%
\providecommand \bibinfo  [0]{\@secondoftwo}%
\providecommand \bibfield  [0]{\@secondoftwo}%
\providecommand \translation [1]{[#1]}%
\providecommand \BibitemOpen [0]{}%
\providecommand \bibitemStop [0]{}%
\providecommand \bibitemNoStop [0]{.\EOS\space}%
\providecommand \EOS [0]{\spacefactor3000\relax}%
\providecommand \BibitemShut  [1]{\csname bibitem#1\endcsname}%
\let\auto@bib@innerbib\@empty
\bibitem [{\citenamefont {Jackson}(2021)}]{jackson2021classical}%
  \BibitemOpen
  \bibfield  {author} {\bibinfo {author} {\bibfnamefont {J.~D.}\ \bibnamefont {Jackson}},\ }\href@noop {} {\emph {\bibinfo {title} {Classical electrodynamics}}}\ (\bibinfo  {publisher} {John Wiley \& Sons},\ \bibinfo {year} {2021})\BibitemShut {NoStop}%
\bibitem [{\citenamefont {Born}\ and\ \citenamefont {Wolf}(1999)}]{born19997th}%
  \BibitemOpen
  \bibfield  {author} {\bibinfo {author} {\bibfnamefont {M.}~\bibnamefont {Born}}\ and\ \bibinfo {author} {\bibfnamefont {E.}~\bibnamefont {Wolf}},\ }\href@noop {} {\bibinfo {title} {7th, principles of optics}} (\bibinfo {year} {1999})\BibitemShut {NoStop}%
\bibitem [{\citenamefont {Silverman}(2008)}]{silverman2008chiral}%
  \BibitemOpen
  \bibfield  {author} {\bibinfo {author} {\bibfnamefont {M.~P.}\ \bibnamefont {Silverman}},\ }\bibfield  {title} {\bibinfo {title} {Chiral asymmetry: The quantum physics of handedness},\ }\href@noop {} {\bibfield  {journal} {\bibinfo  {journal} {Quantum Superposition: Counterintuitive Consequences of Coherence, Entanglement, and Interference}\ ,\ \bibinfo {pages} {257}} (\bibinfo {year} {2008})}\BibitemShut {NoStop}%
\bibitem [{\citenamefont {Silverman}(1985)}]{silverman1985specular}%
  \BibitemOpen
  \bibfield  {author} {\bibinfo {author} {\bibfnamefont {M.}~\bibnamefont {Silverman}},\ }\bibfield  {title} {\bibinfo {title} {Specular light scattering from a chiral medium: unambiguous test of gyrotropic constitutive relations},\ }\href@noop {} {\bibfield  {journal} {\bibinfo  {journal} {Lettere al Nuovo Cimento}\ }\textbf {\bibinfo {volume} {43}},\ \bibinfo {pages} {378} (\bibinfo {year} {1985})}\BibitemShut {NoStop}%
\bibitem [{\citenamefont {Silverman}\ and\ \citenamefont {Sohn}(1986)}]{silverman1986effects}%
  \BibitemOpen
  \bibfield  {author} {\bibinfo {author} {\bibfnamefont {M.}~\bibnamefont {Silverman}}\ and\ \bibinfo {author} {\bibfnamefont {R.}~\bibnamefont {Sohn}},\ }\bibfield  {title} {\bibinfo {title} {Effects of circular birefringence on light propagation and reflection},\ }\href@noop {} {\bibfield  {journal} {\bibinfo  {journal} {American Journal of Physics}\ }\textbf {\bibinfo {volume} {54}},\ \bibinfo {pages} {69} (\bibinfo {year} {1986})}\BibitemShut {NoStop}%
\bibitem [{\citenamefont {Silverman}(1986)}]{silverman1986reflection}%
  \BibitemOpen
  \bibfield  {author} {\bibinfo {author} {\bibfnamefont {M.}~\bibnamefont {Silverman}},\ }\bibfield  {title} {\bibinfo {title} {Reflection and refraction at the surface of a chiral medium: comparison of gyrotropic constitutive relations invariant or noninvariant under a duality transformation},\ }\href@noop {} {\bibfield  {journal} {\bibinfo  {journal} {Journal of the Optical Society of America A}\ }\textbf {\bibinfo {volume} {3}},\ \bibinfo {pages} {830} (\bibinfo {year} {1986})}\BibitemShut {NoStop}%
\bibitem [{\citenamefont {Silverman}\ \emph {et~al.}(1988)\citenamefont {Silverman}, \citenamefont {Ritchie}, \citenamefont {Cushman},\ and\ \citenamefont {Fisher}}]{silverman1988experimental}%
  \BibitemOpen
  \bibfield  {author} {\bibinfo {author} {\bibfnamefont {M.}~\bibnamefont {Silverman}}, \bibinfo {author} {\bibfnamefont {N.}~\bibnamefont {Ritchie}}, \bibinfo {author} {\bibfnamefont {G.}~\bibnamefont {Cushman}},\ and\ \bibinfo {author} {\bibfnamefont {B.}~\bibnamefont {Fisher}},\ }\bibfield  {title} {\bibinfo {title} {Experimental configurations using optical phase modulation to measure chiral asymmetries in light specularly reflected from a naturally gyrotropic medium},\ }\href@noop {} {\bibfield  {journal} {\bibinfo  {journal} {Journal of the Optical Society of America A}\ }\textbf {\bibinfo {volume} {5}},\ \bibinfo {pages} {1852} (\bibinfo {year} {1988})}\BibitemShut {NoStop}%
\bibitem [{\citenamefont {Silverman}\ \emph {et~al.}(1992)\citenamefont {Silverman}, \citenamefont {Badoz},\ and\ \citenamefont {Briat}}]{silverman1992chiral}%
  \BibitemOpen
  \bibfield  {author} {\bibinfo {author} {\bibfnamefont {M.}~\bibnamefont {Silverman}}, \bibinfo {author} {\bibfnamefont {J.}~\bibnamefont {Badoz}},\ and\ \bibinfo {author} {\bibfnamefont {B.}~\bibnamefont {Briat}},\ }\bibfield  {title} {\bibinfo {title} {Chiral reflection from a naturally optically active medium},\ }\href@noop {} {\bibfield  {journal} {\bibinfo  {journal} {Optics letters}\ }\textbf {\bibinfo {volume} {17}},\ \bibinfo {pages} {886} (\bibinfo {year} {1992})}\BibitemShut {NoStop}%
\bibitem [{\citenamefont {Silverman}\ and\ \citenamefont {Badoz}(1990)}]{silverman1990light}%
  \BibitemOpen
  \bibfield  {author} {\bibinfo {author} {\bibfnamefont {M.}~\bibnamefont {Silverman}}\ and\ \bibinfo {author} {\bibfnamefont {J.}~\bibnamefont {Badoz}},\ }\bibfield  {title} {\bibinfo {title} {Light reflection from a naturally optically active birefringent medium},\ }\href@noop {} {\bibfield  {journal} {\bibinfo  {journal} {Journal of the Optical Society of America A}\ }\textbf {\bibinfo {volume} {7}},\ \bibinfo {pages} {1163} (\bibinfo {year} {1990})}\BibitemShut {NoStop}%
\bibitem [{\citenamefont {Silverman}\ and\ \citenamefont {Badoz}(1989)}]{silverman1989large}%
  \BibitemOpen
  \bibfield  {author} {\bibinfo {author} {\bibfnamefont {M.}~\bibnamefont {Silverman}}\ and\ \bibinfo {author} {\bibfnamefont {J.}~\bibnamefont {Badoz}},\ }\bibfield  {title} {\bibinfo {title} {Large enhancement of chiral asymmetry in light reflection near critical angle},\ }\href@noop {} {\bibfield  {journal} {\bibinfo  {journal} {Optics communications}\ }\textbf {\bibinfo {volume} {74}},\ \bibinfo {pages} {129} (\bibinfo {year} {1989})}\BibitemShut {NoStop}%
\bibitem [{\citenamefont {Silverman}(1989)}]{silverman1989differential}%
  \BibitemOpen
  \bibfield  {author} {\bibinfo {author} {\bibfnamefont {M.}~\bibnamefont {Silverman}},\ }\bibfield  {title} {\bibinfo {title} {Differential amplification of circularly polarised light by enhanced internal reflection from an active chiral medium},\ }\href@noop {} {\bibfield  {journal} {\bibinfo  {journal} {Optics communications}\ }\textbf {\bibinfo {volume} {74}},\ \bibinfo {pages} {134} (\bibinfo {year} {1989})}\BibitemShut {NoStop}%
\bibitem [{\citenamefont {Silverman}\ and\ \citenamefont {Badoz}(1992)}]{silverman1992multiple}%
  \BibitemOpen
  \bibfield  {author} {\bibinfo {author} {\bibfnamefont {M.}~\bibnamefont {Silverman}}\ and\ \bibinfo {author} {\bibfnamefont {J.}~\bibnamefont {Badoz}},\ }\bibfield  {title} {\bibinfo {title} {Multiple reflection from isotropic chiral media and the enhancement of chiral asymmetry},\ }\href@noop {} {\bibfield  {journal} {\bibinfo  {journal} {Journal of electromagnetic waves and applications}\ }\textbf {\bibinfo {volume} {6}},\ \bibinfo {pages} {587} (\bibinfo {year} {1992})}\BibitemShut {NoStop}%
\bibitem [{\citenamefont {Ghosh}\ and\ \citenamefont {Fischer}(2006)}]{ghosh2006chiral}%
  \BibitemOpen
  \bibfield  {author} {\bibinfo {author} {\bibfnamefont {A.}~\bibnamefont {Ghosh}}\ and\ \bibinfo {author} {\bibfnamefont {P.}~\bibnamefont {Fischer}},\ }\bibfield  {title} {\bibinfo {title} {Chiral molecules split light: reflection and refraction in a chiral liquid},\ }\href@noop {} {\bibfield  {journal} {\bibinfo  {journal} {Physical review letters}\ }\textbf {\bibinfo {volume} {97}},\ \bibinfo {pages} {173002} (\bibinfo {year} {2006})}\BibitemShut {NoStop}%
\bibitem [{\citenamefont {Debnath}\ and\ \citenamefont {Viswanathan}(2020)}]{debnath2020generalized}%
  \BibitemOpen
  \bibfield  {author} {\bibinfo {author} {\bibfnamefont {A.}~\bibnamefont {Debnath}}\ and\ \bibinfo {author} {\bibfnamefont {N.~K.}\ \bibnamefont {Viswanathan}},\ }\bibfield  {title} {\bibinfo {title} {Generalized matrix transformation formalism for reflection and transmission of complex optical waves at a plane dielectric interface},\ }\href@noop {} {\bibfield  {journal} {\bibinfo  {journal} {Journal of the Optical Society of America A}\ }\textbf {\bibinfo {volume} {37}},\ \bibinfo {pages} {1971} (\bibinfo {year} {2020})}\BibitemShut {NoStop}%
\bibitem [{\citenamefont {Debnath}\ and\ \citenamefont {Viswanathan}(2021)}]{debnath2021generic}%
  \BibitemOpen
  \bibfield  {author} {\bibinfo {author} {\bibfnamefont {A.}~\bibnamefont {Debnath}}\ and\ \bibinfo {author} {\bibfnamefont {N.~K.}\ \bibnamefont {Viswanathan}},\ }\bibfield  {title} {\bibinfo {title} {Generic optical singularities in brewster-reflected postparaxial beam fields},\ }\href@noop {} {\bibfield  {journal} {\bibinfo  {journal} {Physical Review A}\ }\textbf {\bibinfo {volume} {103}},\ \bibinfo {pages} {013510} (\bibinfo {year} {2021})}\BibitemShut {NoStop}%
\bibitem [{\citenamefont {Debnath}\ and\ \citenamefont {Viswanathan}(2022)}]{debnath2022generic}%
  \BibitemOpen
  \bibfield  {author} {\bibinfo {author} {\bibfnamefont {A.}~\bibnamefont {Debnath}}\ and\ \bibinfo {author} {\bibfnamefont {N.~K.}\ \bibnamefont {Viswanathan}},\ }\bibfield  {title} {\bibinfo {title} {Generic optical singularities and beam-field phenomena due to general paraxial beam reflection at a plane dielectric interface},\ }\href@noop {} {\bibfield  {journal} {\bibinfo  {journal} {Physical Review A}\ }\textbf {\bibinfo {volume} {106}},\ \bibinfo {pages} {013522} (\bibinfo {year} {2022})}\BibitemShut {NoStop}%
\bibitem [{\citenamefont {Debnath}\ \emph {et~al.}(2023)\citenamefont {Debnath}, \citenamefont {Kumar}, \citenamefont {Baishya},\ and\ \citenamefont {Viswanathan}}]{debnath2023optical}%
  \BibitemOpen
  \bibfield  {author} {\bibinfo {author} {\bibfnamefont {A.}~\bibnamefont {Debnath}}, \bibinfo {author} {\bibfnamefont {N.}~\bibnamefont {Kumar}}, \bibinfo {author} {\bibfnamefont {U.}~\bibnamefont {Baishya}},\ and\ \bibinfo {author} {\bibfnamefont {N.~K.}\ \bibnamefont {Viswanathan}},\ }\bibfield  {title} {\bibinfo {title} {Optical singularity dynamics and spin-orbit interaction due to a normal-incident optical beam reflected at a plane dielectric interface},\ }\href@noop {} {\bibfield  {journal} {\bibinfo  {journal} {Physical Review A}\ }\textbf {\bibinfo {volume} {107}},\ \bibinfo {pages} {013522} (\bibinfo {year} {2023})}\BibitemShut {NoStop}%
\bibitem [{\citenamefont {Kumar}\ \emph {et~al.}(2024)\citenamefont {Kumar}, \citenamefont {Debnath},\ and\ \citenamefont {Viswanathan}}]{kumar2024complex}%
  \BibitemOpen
  \bibfield  {author} {\bibinfo {author} {\bibfnamefont {N.}~\bibnamefont {Kumar}}, \bibinfo {author} {\bibfnamefont {A.}~\bibnamefont {Debnath}},\ and\ \bibinfo {author} {\bibfnamefont {N.~K.}\ \bibnamefont {Viswanathan}},\ }\bibfield  {title} {\bibinfo {title} {Complex far fields and optical singularities due to propagation beyond tight focusing: combined effects of wavefront curvature and aperture diffraction},\ }\href@noop {} {\bibfield  {journal} {\bibinfo  {journal} {Journal of Optics}\ }\textbf {\bibinfo {volume} {26}},\ \bibinfo {pages} {045604} (\bibinfo {year} {2024})}\BibitemShut {NoStop}%
\bibitem [{\citenamefont {Richards}\ and\ \citenamefont {Wolf}(1959)}]{richards1959electromagnetic}%
  \BibitemOpen
  \bibfield  {author} {\bibinfo {author} {\bibfnamefont {B.}~\bibnamefont {Richards}}\ and\ \bibinfo {author} {\bibfnamefont {E.}~\bibnamefont {Wolf}},\ }\bibfield  {title} {\bibinfo {title} {Electromagnetic diffraction in optical systems, ii. structure of the image field in an aplanatic system},\ }\href@noop {} {\bibfield  {journal} {\bibinfo  {journal} {Proceedings of the Royal Society of London. Series A. Mathematical and Physical Sciences}\ }\textbf {\bibinfo {volume} {253}},\ \bibinfo {pages} {358} (\bibinfo {year} {1959})}\BibitemShut {NoStop}%
\bibitem [{\citenamefont {Novotny}\ and\ \citenamefont {Hecht}(2012)}]{novotny2012principles}%
  \BibitemOpen
  \bibfield  {author} {\bibinfo {author} {\bibfnamefont {L.}~\bibnamefont {Novotny}}\ and\ \bibinfo {author} {\bibfnamefont {B.}~\bibnamefont {Hecht}},\ }\href@noop {} {\emph {\bibinfo {title} {Principles of nano-optics}}}\ (\bibinfo  {publisher} {Cambridge university press},\ \bibinfo {year} {2012})\BibitemShut {NoStop}%
\bibitem [{\citenamefont {Lekner}(1996)}]{lekner1996optical}%
  \BibitemOpen
  \bibfield  {author} {\bibinfo {author} {\bibfnamefont {J.}~\bibnamefont {Lekner}},\ }\bibfield  {title} {\bibinfo {title} {Optical properties of isotropic chiral media},\ }\href@noop {} {\bibfield  {journal} {\bibinfo  {journal} {Pure and Applied Optics: Journal of the European Optical Society Part A}\ }\textbf {\bibinfo {volume} {5}},\ \bibinfo {pages} {417} (\bibinfo {year} {1996})}\BibitemShut {NoStop}%
\bibitem [{\citenamefont {Condon}(1937)}]{condon1937theories}%
  \BibitemOpen
  \bibfield  {author} {\bibinfo {author} {\bibfnamefont {E.~U.}\ \bibnamefont {Condon}},\ }\bibfield  {title} {\bibinfo {title} {Theories of optical rotatory power},\ }\href@noop {} {\bibfield  {journal} {\bibinfo  {journal} {Reviews of modern physics}\ }\textbf {\bibinfo {volume} {9}},\ \bibinfo {pages} {432} (\bibinfo {year} {1937})}\BibitemShut {NoStop}%
\bibitem [{\citenamefont {Tellegen}(1948)}]{tellegen1948gyrator}%
  \BibitemOpen
  \bibfield  {author} {\bibinfo {author} {\bibfnamefont {B.~D.}\ \bibnamefont {Tellegen}},\ }\bibfield  {title} {\bibinfo {title} {The gyrator, a new electric network element},\ }\href@noop {} {\bibfield  {journal} {\bibinfo  {journal} {Philips Res. Rep}\ }\textbf {\bibinfo {volume} {3}},\ \bibinfo {pages} {81} (\bibinfo {year} {1948})}\BibitemShut {NoStop}%
\bibitem [{\citenamefont {Goldstein}(2017)}]{goldstein2017polarized}%
  \BibitemOpen
  \bibfield  {author} {\bibinfo {author} {\bibfnamefont {D.~H.}\ \bibnamefont {Goldstein}},\ }\href@noop {} {\emph {\bibinfo {title} {Polarized light}}}\ (\bibinfo  {publisher} {CRC press},\ \bibinfo {year} {2017})\BibitemShut {NoStop}%
\bibitem [{\citenamefont {Baishya}\ and\ \citenamefont {Viswanathan}(2023)}]{baishya2023measurement}%
  \BibitemOpen
  \bibfield  {author} {\bibinfo {author} {\bibfnamefont {U.}~\bibnamefont {Baishya}}\ and\ \bibinfo {author} {\bibfnamefont {N.~K.}\ \bibnamefont {Viswanathan}},\ }\bibfield  {title} {\bibinfo {title} {Measurement of surface chirality at near-normal incidence},\ }\href@noop {} {\bibfield  {journal} {\bibinfo  {journal} {Applied Physics Letters}\ }\textbf {\bibinfo {volume} {122}},\ \bibinfo {pages} {261102} (\bibinfo {year} {2023})}\BibitemShut {NoStop}%
\bibitem [{\citenamefont {Volyar}\ \emph {et~al.}(2007)\citenamefont {Volyar}, \citenamefont {Rubass}, \citenamefont {Shvedov}, \citenamefont {Fadeyeva},\ and\ \citenamefont {Kotlyarov}}]{volyar2007optical}%
  \BibitemOpen
  \bibfield  {author} {\bibinfo {author} {\bibfnamefont {A.}~\bibnamefont {Volyar}}, \bibinfo {author} {\bibfnamefont {A.}~\bibnamefont {Rubass}}, \bibinfo {author} {\bibfnamefont {V.}~\bibnamefont {Shvedov}}, \bibinfo {author} {\bibfnamefont {T.}~\bibnamefont {Fadeyeva}},\ and\ \bibinfo {author} {\bibfnamefont {K.}~\bibnamefont {Kotlyarov}},\ }\bibfield  {title} {\bibinfo {title} {Optical vortices and airy’spiral in chiral crystals},\ }\href@noop {} {\bibfield  {journal} {\bibinfo  {journal} {Ukr. J. Phys. Opt}\ }\textbf {\bibinfo {volume} {8}},\ \bibinfo {pages} {166} (\bibinfo {year} {2007})}\BibitemShut {NoStop}%
\bibitem [{\citenamefont {Chipman}\ \emph {et~al.}(2018)\citenamefont {Chipman}, \citenamefont {Lam},\ and\ \citenamefont {Young}}]{chipman2018polarized}%
  \BibitemOpen
  \bibfield  {author} {\bibinfo {author} {\bibfnamefont {R.}~\bibnamefont {Chipman}}, \bibinfo {author} {\bibfnamefont {W.~S.~T.}\ \bibnamefont {Lam}},\ and\ \bibinfo {author} {\bibfnamefont {G.}~\bibnamefont {Young}},\ }\href@noop {} {\emph {\bibinfo {title} {Polarized light and optical systems}}}\ (\bibinfo  {publisher} {CRC press},\ \bibinfo {year} {2018})\BibitemShut {NoStop}%
\end{thebibliography}%

\end{document}